# A New Approach to Real Time Impulsive Sound Detection for Surveillance Applications

Yüksel Arslan

*Abstract*—Most of the surveillance systems for public safety are solely based on one or more video cameras. These camera systems have some drawbacks such that they have poor performance in adverse weather conditions or during night time. Therefore most of the time, some other sensors should accompany to video cameras. Although audio surveillance is in its early stage, there has been considerable amount of work in this area in the last decade. In this paper we make a review of impulsive sound detection algorithms. Sounds from dangerous events such as gunshots, explosions, human screaming can be classified as impulsive sounds, so this paper reviews all impulsive sound detection algorithms along with impulsive noise detection algorithms although they progress in their own path. These dangerous sound events have no other detection means except audio.

We try to adapt some algorithms used in impulsive noise detection to the area of impulsive sound detection. Tests show that Warped Linear Prediction (WLP) can be used for impulsive sound detection.

## I. INTRODUCTION

Environmental Sound Recognition (ESR) can be used for many different kinds of purposes. Dangerous sound event recognition is a subset of ESR. A typical ESR system has the following parts: Microphone input and digitization, detection, feature extraction and recognition. This system can be run in real-time or offline after recording. Generally a surveillance system for hazardous sound event detection works alone or incorporates a video surveillance system.

Figure 1 is showing a typical ESR system [1]. This system can be used to detect and recognize dangerous audio events. Then, the data flow is as follows. Environmental sounds are captured and digitized by a microphone input and a sound card. Sampling rate and quantization bit depth are adjusted here. Sampling rate can affect the performance, because each sample is handled at the detection stage separately. We should also consider the buffering size here which will affect the performance later. Detection stage comes after the input and digitization of the sound input. Detection stage detects the impulsive sounds. Impulsive sounds are the result of some dangerous events such as gunshots, explosions, human screams etc. or they may be from other sources such as thunder, helicopter [2]. In detection stage just framing (windowing) or first framing then sub-framing techniques can be applied. Some algorithms inspect each sample alone to detect if it is an impulse or not without framing. Frame size has a big impact on the detection of impulsive sounds and later in the recognition stage. Detection algorithms always work in the system. If an impulsive sound is detected then the detected frame is passed to recognition stage. For real-time constraints, a dangerous sound event detection algorithm must run near real-time. Another important thing we should consider in detection stage is the threshold value. Threshold is applied on the output of the algorithm for decision. The threshold value should not be very low to overload recognition stage or it should not be very high to miss dangerous sound events. The main objective of the detection stage is to decide on frames which consist an impulsive sound and then to pass these frames to the recognition stage.

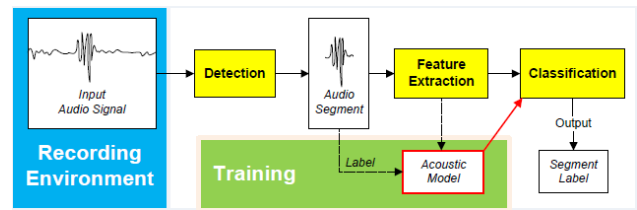

Figure 1 A typical ESR system [1]

On the other hand in some systems detection stage can be omitted. In section II we will see some algorithms which we classified as detection by recognition. ESR systems using these algorithms don't need the detection stage [3][4].

After detection of impulsive sounds, recognition stage takes the frame and decides if it is from a dangerous event or not and the class of the event.

In this paper we will explore the algorithms used at the detection stage. We will make a novel contribution such that we will also study the usability of impulsive noise detection algorithms at this stage.

This paper is organized as follows: In Section II a novel taxonomy of impulsive sound/noise detection algorithms is given and explained. In Section III impulsive sound detection methods and algorithms for each method are explained. In Section IV some selected algorithms from impulsive noise detection group are implemented for detection of impulsive sounds. Performance comparison is made based on miss detections and false positives. In Section V we will explain our contributions, conclusion and future work.

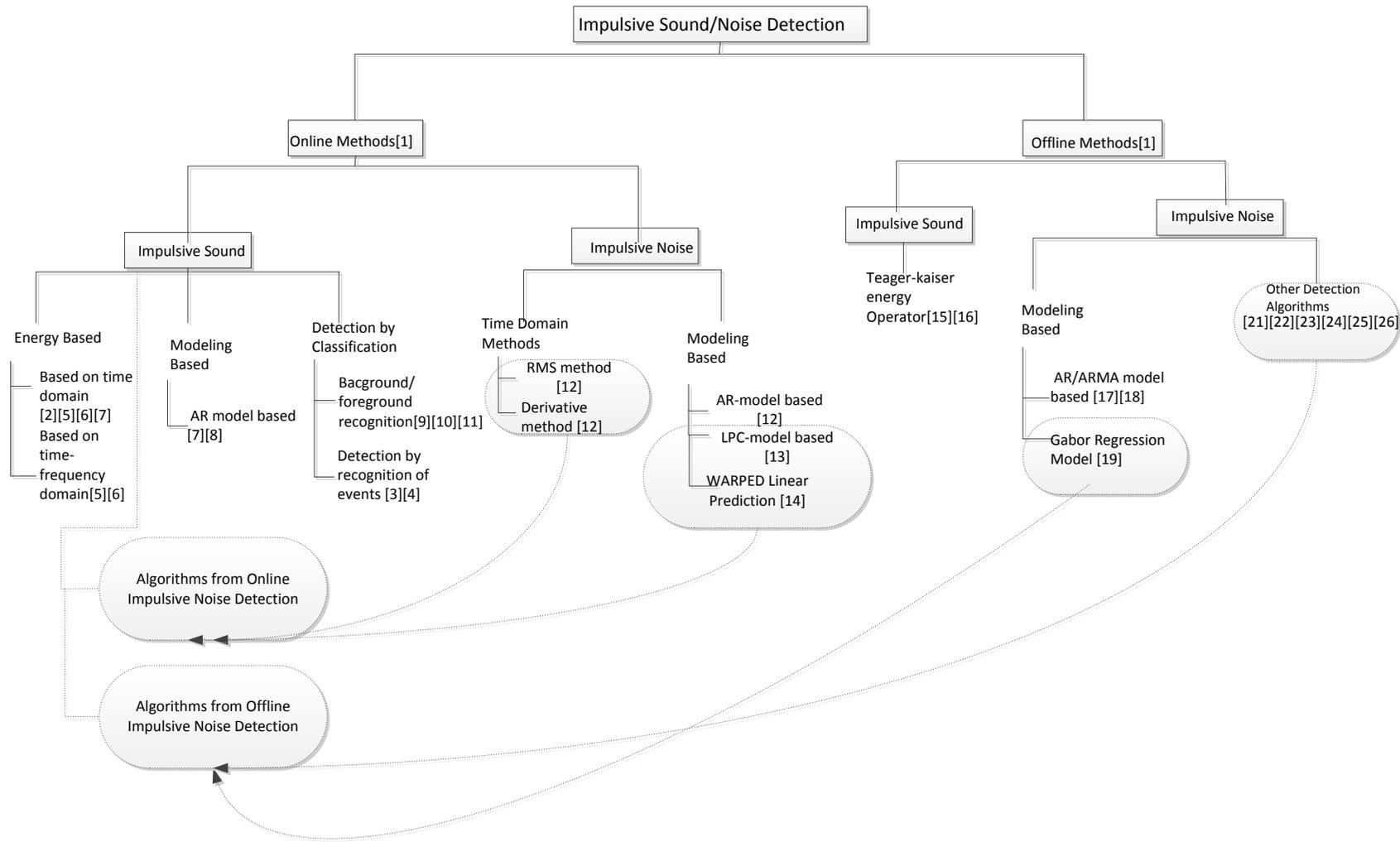

**Figure 2 Taxonomy of Impulsive Sound/Noise Detection Algorithms and Our New Approach for Impulsive Sound Detection**

## II. IMPULSIVE SOUND/NOISE DETECTION ALGORITHMS

### A. Classification of Impulsive Sound/Noise Detection Algorithms:

In Figure 2 a classification of impulsive sound/noise detection algorithms is shown. Impulsive sound/noise detection algorithms can be grouped into two at the top, regarding the detection time. Online or real-time algorithms detect the impulsive sound/noise during the dangerous event happens. Offline algorithms detect it from the sound recordings of the event after the event has happened [1].

Furthermore the detection algorithms can be divided into two subclasses, as impulsive sound and impulsive noise detection algorithms. Impulsive sounds are the results of an acoustic event, they possess high energy and they last longer than impulsive noises. They have certain patterns so they can be classified by machine learning algorithms. Impulsive noise occurs in analog transmission media such as communications, telephony, and radio broadcast caused by electromagnetic disturbances and atmospheric noise. Impulsive noise results also from physical damages to the storage media e.g. degraded gramophone recordings [13]. Impulsive disturbances or clicks can be described as short time (1 to 200 samples at 44100Hz sampling rate) local discontinuities [19].

Hazardous events such as gunshots, explosions, glass breaks, door slams generate impulsive sounds. Audio surveillance applications which are based on impulsive sound detection and the applications which enhance the sound quality of recording media and communication try to detect impulsive sound and impulsive noise respectively using their own algorithms.

It should be noted here that there is no restriction to use online impulsive sound or noise detection algorithms for offline applications.

### B. Online Impulsive Sound Detection Algorithms:

#### a. Algorithms based on signal energy:

The most common parameter that can be used to discriminate an impulsive sound from the background is its energy. The sounds that we are interested in such as gunshots, door slams, screams cause an abrupt change in audio volume depending on the typical auditory scenarios. Following this principle, some works proposed to segment the audio stream into fixed-length windows and discard all the windows whose energy is below a predefined threshold [27]. In [2], it is proposed three methods using above principle. A real-time environmental sound recognition system is implemented in [28] for Android operating system. In this system an energy detector which is based on one of the algorithms defined in [2] is used and reported that it was fast and efficient. For robust localization of impulsive sounds, an energy detector is used first to detect impulsive and non-impulsive sound sources in [6]. Besides these algorithms which run in time domain there are other type of algorithms using this principle running in time-frequency domain [5][7].

Energy based algorithms are generally fast compared to the others but there are problems related to these algorithms, described below:

**Threshold adjusting:** Actually this problem is related to most of the detection algorithms. Impulsive sound decision is given based on a threshold value. If the energy of current window is above the threshold value then the window contains impulsive sound. The threshold value can be a fixed value or adaptive. This value should be set so that the system would not miss impulsive sounds. We call these miss detections. If the threshold level is too high then miss detections occur. On the contrary if the threshold level is too low then we detect other sounds we are not interested in. It is called false positives. This puts an extra load on the recognition stage and the system slows down. To determine the threshold value, fixed threshold value [29], set of thresholds [30] and adaptive thresholding [2] was proposed.

**Windowing:** The algorithms described above based on energy, segments the signal into fixed length blocks called windowing. This fact may chop a significant audio event into two adjacent blocks so making it more difficult for the subsequent processing stages [27]. For example before the recognition stage features must be extracted from the complete sound of the acoustic event for correct recognition.

The other problem regarding windowing is that every impulsive sound has different length. So, selecting one window size doesn't match for all the impulsive sounds.

#### b. Algorithms based on sound modeling:

In [8][9], to detect impulsive sounds from the muzzle blasts of gunshots, Auto Regressive (AR) model is used. Although AR model operates on time domain data, it is capable of encapsulating both spectral and energy characteristics of time series. AR model doesn't need any segmentation of the signal. To accurately model the background signal, non-stationary AR model is proposed as the most appropriate model. Least Mean Square (LMS) is used to find the parameters of the non-stationary AR model [8][9].

#### c. Detection by classification:

We can divide the algorithms used here into two categories. The first category combines the methods that classify the sounds in the environment as background or foreground, later the algorithm tries to recognize foreground sounds which are called abnormal sound events. In this first category, it is generally assumed that there is no knowledge of foreground (abnormal) sounds hence it is not possible to train the algorithm in advance. The algorithm learns the background generally online, and then it can detect the abnormal events in the environment. In [10], a Gaussian Mixture Model (GMM) is proposed for modelling the background sounds. This model updates itself incrementally when the new sounds come. The proposed model is a hybrid one which detects the foreground suspicious activity online and then this suspicious activity is classified after inspection as usual or unusual activity.

In [11], the Time-Adaptive mixture of Gaussians method, well-known in the video surveillance context, which aims at discovering the deviance of a signal from the expected behavior in an on-line fashion, is used. Windowed audio signal is transformed to frequency domain and the energy of different sub bands are modeled by different GMMs. This model is also adaptive to background sounds and can detect abnormal sounds.

In [12], a technique that uses multiple GMMs for different level of sound events is explained. The basic idea of the multi-stage GMM is to model the majority of the training samples using the first GMM, and then model the rest of the samples using the second GMM. The training stage goes until usual sounds are classified by the end of the GMMs. If a sound signal comes with likelihood less than the last GMM's, then this sound is classified as abnormal.

The methods in the second category of detection by classification can classify the sounds into predefined and trained classes. For real-time applications the computation time must be considered for this category. In [3], it is proposed that the observed sound signal is composed of signal and noise and the signal and noise subspaces are orthogonal to each other. As a consequence of this, noise effect is reduced by projecting the observation vector onto signal subspace. During the training phase Principle Component Analysis (PCA) technique is used to create event subspaces. To construct feature vectors Short Time Fourier Transform (STFT) is taken and PCA is calculated by including 90% of the signal energy. These features are used with Support Vector Machine (SVM), GMM and Deep Belief Network (DBN) to recognize events such as scream, glass break, and siren. It is shown that the proposed features outperform conventional features such as Mel Frequency Cepstral Coefficients (MFCC) when used with the same algorithm.

In [4], one-class SVMs are proposed for audio surveillance systems. This paper explains a feature vector selection algorithm and constructing a fixed length feature vector from different duration signals. Lastly it is explained that wavelets are appropriate as features for short duration and non-stationary signals.

C. *Online Impulsive Noise Detection Algorithms:*

    a. *Time domain methods:*

There are two algorithms in this category. Derivative method uses the definition of impulsive sound which is sudden change between successive samples. So, derivative method is based on tracking this change. Fourth derivative gives more accurate results to detect the impulsive noises. Also in [13] an adaptive threshold is applied on the pre-determined number of past derivative values.

The other method is the Root Mean Square (RMS). The root mean square of the windowed samples is calculated [13].

    b. *Sound modeling based:*

In [13], previously explained AR method used for online impulsive sound detection is used first to predict the incoming samples. AR order is taken as 10. The AR prediction errors are found and then to select the impulsive samples previously explained derivative method is used. The difference between this algorithm and the previous one is that while this algorithm accepts the signal as stationary, the other accepts non-stationary. This difference leads to calculation of model parameters differently.

In [14], it is declared that the Linear Predictive Coding (LPC) is a good way to model speech signals whereas it is not good for impulsive disturbances. LPC order is taken as 20. After detecting impulsive noises, an interpolation algorithm is proposed to replace noisy samples. It is reported that it gives successful results on numerous noisy recordings.

It should be noted that there is little difference between AR model and LPC model. AR and LPC model coefficients are the same if you take the same order. But, the interpretation is different.

In [15], it is explained how Warped Linear Prediction (WLP) is used to detect impulsive noises. WLP operates in the frequency domain and to detect the impulsive noises changes the prediction gain. A comparison is made in [15] between LPC and WLP and it is shown that WLP decreases miss and false detections when the parameters are chosen correctly.

D. *Offline Impulsive Sound Detection Algorithms:*

Offline impulsive sound detection may be useless, because the aim of detection is to respond to real time events immediately. So there are just a few researches on offline impulsive sound detection. These impulsive sounds are not from dangerous audio events.

In [16], Teager-Kaiser (TK) operator is used to detect the clicks of sperm whales in large recordings. Selection of clicks is based on applying a statistical measurement on TK operator output. As a result 94.5% correct detection was achieved.

In [17], to distinguish prerecorded sounds of six species of odontocetes TK operator was used. Sound recordings are windowed and converted to frequency domain, and then TK operator is applied on the high frequency components. Detected windows are recognized using GMM.

E. *Offline Impulsive Noise Detection Algorithms:*

    a. *Algorithms based on sound modeling:*

As in online impulsive sound detection, AR algorithm is used to model the background sound signal. In [18], AR method is used to detect impulsive noises such as clicks, bursts or scratches. The method is tested on manually distorted several types of sounds (classical, jazz, vocal etc.) and shows good performance.

In [19], detailed theoretical background of AR/ARMA model is given and explained how to remove clicks from sound signals. AR model can be extended with ARMA given more accurate localization of clicks.

In [20], Gabor regression model is used to determine the positions of impulsive noises. Just as Fourier Transform is a representation for stationary signals, Gabor representations are appropriate for slowly time-varying signals [20].

    b. *Other detection algorithms:*

There are many different impulsive noise detection algorithms in literature that can't be grouped into small number of groups. Here we mention just the most recent ones. In [21], five different outlier detection algorithms and an outlier detection algorithm which utilizes prediction algorithms for univariate time series are proposed. Proposed outlier detection algorithm is tested with twelve different forecasting models. The proposed Absolute Predictive Deviation (APD) is given the best result when used with the Autoregressive Integrated Moving Average (ARIMA) model.

In [22], a bidirectional processing is proposed. In this new approach, noise pulses are localized by combining the results of forward-time and backward-time signal analysis. For forward-time and backward-time signal analysis AR model is used.

In [23], the proposed semicausal and noncausal outlier detection techniques are demonstrated on stereo gramophone recordings.

In [24], detection of noise pulses in stereo recording is achieved by using vector AR model and Kalman Filtering.

In [25], it is shown that semi-causal/non-causal solutions based on joint evaluation of signal prediction errors and leave-one-out signal interpolation errors improve detection results compared to prediction-only based solutions.

In [26], the method called binary time-frequency masking threshold criterion is proposed to detect and remove Gaussian, Supergaussian and impulsive disturbances. The noisy signal is decomposed into sum of sinusoids and a residual. AR model is used on the residual to detect the impulsive noise.

## III. IMPULSIVE SOUND DETECTION METHODS

By using above algorithms to construct an impulsive sound detection and recognition system, the detection stage possibilities are explained here.

The detection only method is seen in Figure -3.

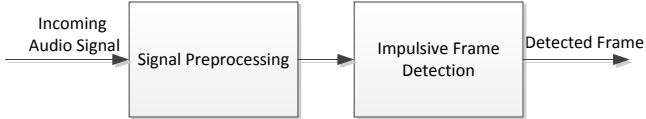

**Figure 3 Impulsive frame detecion method**

In Figure-3 impulsive frame detection block can use one of the algorithms from;

- Energy based algorithms, [2][5][6][7]
- Prediction based algorithms, from online impulsive sound detection and online/offline impulsive noise detection algorithms [8][9][13][14][15][18][19][20]
- Offline impulsive noise detection algorithms presented in [21][22][23][24][25][26]. These can be used in this method if it runs in time limitations of application requirement.

Background/foreground recognition method is seen in Figure-4.

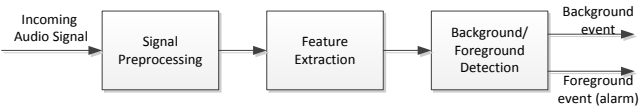

**Figure 4 Background/foreground recognition method**

In this method after preprocessing, a machine learning algorithm can be used to differentiate between background and foreground sounds. The foreground sounds for our application are sounds of interested hazardous events. The application can be trained in advanced or online for required impulsive events. In this method the algorithms defined in papers [10][11][12] can be used.

Impulsive event recognition is shown in Figure-5.

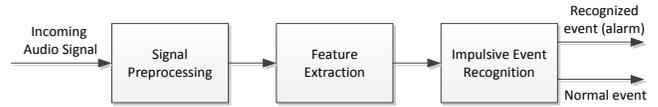

**Figure 5 Impulsive event recognition method**

In this method event recognition block is trained before for required events. It recognizes the events by inspecting features extracted from incoming signal. The algorithms defined in papers [3][4] can be used here.

## IV. IMPLEMENTATIONS AND RESULTS

In this section we will implement some of the impulsive noise detection algorithms which are not tested before namely LPC and WLP algorithms for impulsive sound detection especially considering online detection. We implemented these algorithms because these are firstly implemented for impulsive sound detection purpose. These impulsive noise detection algorithms can be used in the first method described in Section III. We will also compare the performances of these algorithms with an energy base algorithm.

### A. Evaluation Data and Method:

This is a preliminary implementation and the aim is just to see if impulsive noise detection algorithms can be used for online impulsive sound detection. For this purpose we prepared a test sound database from Detection and Classification of Acoustic Scenes and Events (DCASE) 2017 Task-2 database. Task-2 database contains rare sound events which are gunshot, baby cry and glass break. For the tests we used gunshot and glass break sounds. For detection task, these sounds are mixed with 15 different background sounds. Totally there are 250 mixed sounds with a length of 30 sec from each sound class. Each mixture is prepared at one of 3 different event-to-background-ratio (EBR) levels. These levels are -6, 0 and 6. The EBR was defined as ratio of average RMS Error (RMSE) values calculated over the duration of the event and the corresponding background segment on which the event will be mixed respectively. The detailed explanation of the dataset can be found in [31]. These sound clips are loaded to the database developed in [32] for easy usage. The algorithm implemented in [2] that is threshold on the power sequence is used as the base algorithm for our tests, because we saw in previous tests that this algorithm was performing well to detect impulsive sounds. Comparison is made based on miss detections (MD), and false positives (FP). MD are not detected embedded gunshot sounds. Correlate of this is true positives (TP), which means correctly detected embedded gunshot sounds. FP are possible impulses which are not embedded by us. It should be noted here that these FP can actually be impulsive sounds but it is not important for us and it does not show that algorithm is running incorrectly. MD and FP should be as low as possible. Detection tasks can be viewed as involving a tradeoff between FP and MD. Detection Error Tradeoff (DET) [33] curves show the performance of an algorithm by showing the MD and FP rates on the same graph for a set of different threshold values. As the algorithm DET curve closes to the origin performance increases, otherwise it decreases.

### B. Implementations:

#### a. Base algorithm implementation:

The base algorithm implemented for comparison is the algorithm defined in [2] that is threshold on the power sequence. The energy detector is optimum solution if the noise and the signal are uncorrelated zero mean Gaussian random vectors [6]. The base algorithm, first windows incoming samples with a length of 350 samples. Then energy of these frames is calculated. The standard deviation and mean of last 30 energies are calculated. A threshold value is calculated based on this standard deviation and mean. If the incoming frame energy is above this threshold it is an impulse, otherwise it is not impulse.

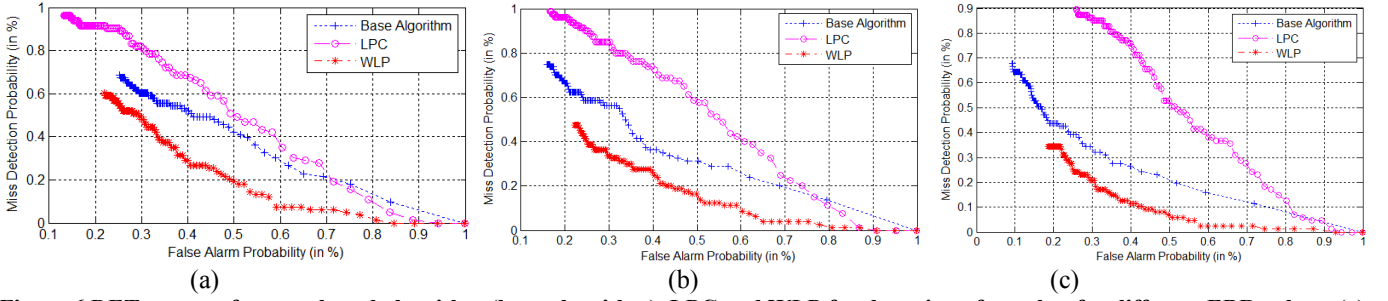

**Figure 6 DET curves of energy based algorithm (base algorithm), LPC and WLP for detection of gunshot for different EBR values. (a) EBR = -6, (b) EBR = 0, (c) EBR = 6**

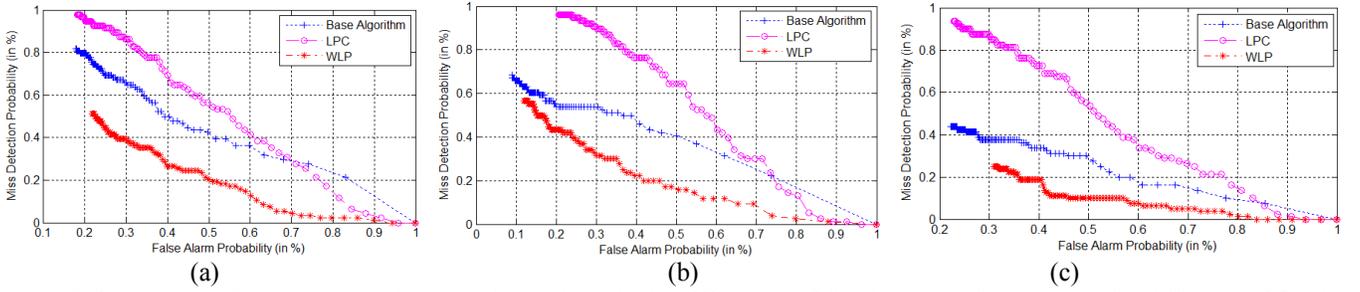

**Figure 7 DET curves of energy based algorithm (base algorithm), LPC and WLP for detection of glass break for different EBR values. (a) EBR = -6, (b) EBR = 0, (c) EBR = 6**

*b. LPC:*

LPC implementation first finds the LPC coefficients by using a prediction order 5. These coefficients can be used for a period of the sound which is assumed to be stationary. Then calculation is redone. Filtering the signal using LPC coefficients an LPC estimate of the signal is found. Difference between the original and the estimate gives the error. By applying a threshold on to the error, impulsive noise is obtained. LPC formulation steps are given as follows:

1- A sound signal modeling by LPC is:

$$x_n[n] = \sum_{i=1}^{N} a_i x_i[n-i] + e_n \quad (1)$$

2- If there is an impulsive sound:
$$y_n = x_n + e_n + d_n \quad (2)$$

3- By passing $y_n$ from a Finite Impulse Response (FIR) filter we can obtain error with impulsive sound. FIR transfer function is:

$$H(z) = 1 + \sum_{i=1}^{N} \bar{a}_i z^{-i} \quad (3)$$

4- We can detect the impulsive sound if the error exceeds a threshold value.

The details can be found in [34].

*c. WLP:*

In WLP the delay elements of the FIR filter in LPC are replaced by first order all-pass filters [15]. Then the delay element is as follows:

$$\tilde{z} = \frac{z^{-1} - \lambda}{1 - \lambda z^{-1}} \quad |\lambda| < 1 \quad (4)$$

By replacing in this new formula just the delay element with $\lambda = -0.7$ and applying the same formulation in LPC, impulsive sounds can be obtained. By applying the same parameters as in LPC, it is seen that WLP increases TP and decreases FP, as described in [15]. We implemented the WLP using the tools given at [35].

*C. Results:*

In this section, WLP and LPC algorithms are implemented and tested. Performances of these algorithms are compared with an energy based algorithm. The results are shown in Figure 6 and Figure 7. It is shown that base algorithm stands between LPC and WLP algorithms. The WLP algorithm outperforms energy based algorithm (base algorithm).

**Table 1 Execution time of algorithms**

| Algorithm | Execution Time(sec) |
|---|---|
| Base | 0.104 |
| LPC | 1.36 |
| WLP | 0.905 |

Execution time of algorithms is shown in Table 1. The time is measured for processing of one 30 sec mixed sound clip. The platform used is a laptop with Intel(R) Core(TM) i5 CPU (2x 2.67 GHz), 4 GB RAM and Matlab 11a. WLP is slower than the base algorithm but it still executes in real-time.

The base algorithm employs windowing. Windowing has several disadvantages. One is the impulsive sound may chop next window, so it may be difficult to determine start and end of the impulsive sound. The second is even if it is in the detected window the start of impulsive sound is not exact, somewhere in the window. This ambiguity in start of impulsive sound will affect the performance of recognition stage and location finder stage if employed. More importantly the window length predetermined for a single dangerous audio event such as gunshot may not be appropriate for other impulsive sound events. The detection performance achieved by a windowing length determined for a gunshot may not be achieved for explosions.

All algorithms classified under impulsive noise detection (Figure 2) work without windowing. They directly detect the start of impulsive sounds. So they don't possess the disadvantages of windowing.

## V. CONCLUSION AND FUTURE WORK

To overcome the limitations of video surveillance systems currently installed, audio sensors with detection and recognition algorithms may be employed. Impulsive sound detection and impulsive noise detection appear as two distinct research areas. We classified the algorithms of these areas (Figure 2). We implemented LPC and WLP algorithms, which are traditionally used for impulsive noise detection, for impulsive sound detection along with energy based algorithm. Detection stage is important for the overall performance of the system. In this stage if there are a lot of miss detections, then the system performance degrades and if there are a lot of false detections then by putting extra load on the recognition stage, it also degrades the performance. So the algorithms used in detection stage are critical.

In our study, the LPC and WLP algorithms are employed, to the best of our knowledge, for the first time for impulsive sound detection. The obtained results show that WLP performs better for impulsive sound detection than the widely used energy based detector. Although WLP is 0.801 sec slower than energy based detector, it can be still used for real time applications.

Future work includes employing these algorithms together with a classification stage in real world surveillance applications.

REFERENCES

[1] J. W. Dennis, "Sound Event Recognition in Unstructured Environment Using Spectogram Image processing", Ph.D. Thesis, Nanyang Technological University, 2014.

[2] A. Dufaux, "Detection and recognition of Impulsive Sound Signals", Ph.D. Thesis, University of Neuchatel, 2001.

[3] S. Park, Y. Lee, D.K. Han and H. Ko, "Subspace projection cepstral coefficients for robust acoustic event detection", in *Proceedings of the IEEE International Conference on Acoustics, Speech, and Signal Processing (ICASSP)*, 2017.

[4] A. Rabaoui, M. Davy, S. Rossignol, Z. Lachiri, N. Ellouze, "Using one-class SVMs and wavelets for audio surveillance systems", *IEEE Trans. Inf. Forensic Security*, 2007.

[5] M. Vacher, D. Istrate, L. Besacier, J. Serignat, E. Castelli, "Life Sounds Extraction and Classification in Noisy Environment", in *Proc. IASTED International Conference on Signal & Image Processing*, 2003.

[6] T. Machmer, A. Swerdlow, K. Kroschel, J. Moragues, at all, "Robust Impulsive Sound Source Localization By Means of an Energy Detector for Temporal Alignment and Pre-Classification", *European Signal Processing Conference*, 17th, 2009.

[7] M. Jorge, S. Arturo, L. Vergara and J. Gosálbez, "Improving detection of acoustic signals by means of a time and frequency multiple energy detector", *IEEE Signal Processing Letters*, vol. 18, n. 8, pp. 458-461, 2011.

[8] Kenneth D. Morton Jr.,"Bayesian Techniques for Adaptive Acoustic Surveillance", Ph.D. Thesis, Duke University, 2010.

[9] Kenneth D. Morton Jr., "Bayesian Detection of Acoustic Muzzle Blasts", *Proc. of SPIE* Vol. 7305 730511-1 2009.

[10] Radhakrishnan, R., Divakaran, A., and Smaragdis, P., "Audio analysis for surveillance applications", in *Applications of Signal Processing to Audio and Acoustics. IEEE Workshop on*, pp. 158 – 161, 2005.

[11] Cristani, M., Bicego, M., and Murino, V., "On-line adaptive background modelling for audio surveillance", in *Proceedings of the 17th International Conference on Pattern Recognition*, vol. 2, pp. 399 – 402, 2004.

[12] Ito, A., Aiba, A., Ito, M., and Makino, S., "Detection of abnormal sound using multi-stage GMM for surveillance microphone", in *Information Assurance and Security,. IAS '09. Fifth International Conference on*, vol. 1, pp. 733 –736, 2009.

[13] lsmo Kauppinen, "Methods for Detecting Impulsive Noise in Speech and Audio Signals", *Proc. 14th Int. Conf. Digit. Signal Process. (DSP2002),* vol. 2, pp. 967-970, 2002.

[14] S. Vaseghi, P. Rayner, "Detection and suppression of impulsive noise in speech communications systems", *Proc. Inst. Elec. Eng. Commun. Speech Vision*, pp. 38-46, Feb 1990.

[15] P. A. A. Esquef, M. Karjalainen, V. Vlimki, "Detection of clicks in audio signals using warped linear prediction", *14th IEEE Int. Conf. Digital Signal Process. (DSP-02)*, 2002.

[16] M. A. Roch, H. Klinck, S. Baumann-Pickering, D. K. Mellinger, S. Qui, "Classification of echolocation clicks from odontocetes in the Southern California Bight", *Journal of the Acoustical Society of America*, 129: 467-475, 2011.

[17] V. Kandia, Y. Stylianou, "Detection of Sperm Whale Clicks Based On the Teager – Kaiser Energy Operator", *Applied Acoustics* 67 pp. 1144–1163, 2006.

[18] L. Oudre, "Automatic detection and removal of impulsive noise in audio signals", *Image Processing On Line*. Preprint. February 2014.

[19] Simon J. Godsill and Peter J.W. Rayner, "Digital Audio Restoration - a statistical model based approach", book 1998

[20] J. Murphy and S. Godsill, "Joint Bayesian removal of impulse and background noise", in *Proceedings of the IEEE International Conference on Acoustics, Speech, and Signal Processing (ICASSP)*, pp. 261-264, 2011.

[21] C.F. Stallmann and A.P. Engelbrecht, "Predictive Noise Detection in Gramophone Records", *International Journal of Signal Processing Systems*, Vol. 4, No. 6, December 2016.

[22] M. Nied´zwiecki and M. Ciołek, "Elimination of impulsive disturbances from archive audio signals using bidirectional processing," *IEEE Trans. Audio, Speech, Lang. Process.*, vol. 21, no. 5, pp. 1046–1059, May 2013.

[23] M. Niedzwiecki and M. Ciolek, "New Semicausal and Noncausal Techniques for Detection of Impulsive Disturbances in Multivariate Signals with Audio Applications", *IEEE Transactions on Signal Processing*, Vol. 65, No:15, August 2017.

[24] M. Nied´zwiecki, M. Ciołek, and K. Cisowski, "Elimination of impulsive disturbances from stereo audio recordings using vector autoregressive modeling and variable-order Kalman filtering," IEEE Trans. Audio, Speech, Lang. Process., vol. 23, no. 6, pp. 970–981, Jun. 2015.

[25] M. Ciołek, M. Nied´zwiecki, "Detection of impulsive disturbances in archive audio signals", in *Proceedings of the IEEE International Conference on Acoustics, Speech, and Signal Processing (ICASSP),2017.*

[26] M. Ruhland, J. Bitzer, M. Brandt, and S. Goetze, "Reduction of Gaussian, supergaussian, and impulsive noise by interpolation of the binary mask residual," *IEEE/ACM Transactions on Audio, Speech, and Language Processing*, vol. 23, no. 10, pp. 1680 – 1691, 2015.

[27] M. Crocco, M. Cristani, A. Trucco, V. Murino, "Audio surveillance: A systematic review", *CoRR*, vol. abs/1409.7787, 2014.

[28] A. Pillos, K. Alghamidi, N. Alzamel, V. Pavlov and S. Machanavajhala, "A Real-Time Environmental Sound Recognition System for The Android OS", *Detection and Classification of Acoustic Scenes and Events"*, 3 Sep. 2016, Hungary.

[29] M. Azlan,, I. Cartwright, N. Jones, T. Quirk, and G. West, "Multimodal monitoring of the aged in their own homes", in *Proceedings of the 3rd International Conference on Smart Homes and Health Telematics (ICOST'05)*, 2005.

[30] A. F. Smeaton and M. McHugh, "Towards event detection in an audio-based sensor network", in *Proc. 3rd Int. Workshop Video Surveill. Sens. Netw. (VSSN), pp. 87-94, Nov. 2005.*

[31] A. Mesaros, T. Heittola, A. Diment, B. Elizalde, A. Shah, E. Vincent, B. Raj, and T. Virtanen, "DCASE 2017 challenge setup: tasks, datasets and baseline system", in *Proceedings of the Detection and Classification of Acoustic Scenes and Events 2017 Workshop (DCASE2017)*, November 2017.

[32] Y.Arslan and H. Canbolat, "A sound database development for environmental sound recognition", *Signal Processing and Communications Applications Conference (SIU),* 25[th], 2017.

[33] A. Martin, G. Doddington, T. Kamm, M. Ordowski, and M. Przybocki, "The DET Curve in Assessment of Detection Task Performance," DTIC Document, Tech. Rep., 1997.

[34] P. P. Vaidyanathan, "The Theory of Linear Prediction", book, 2008.

[35] http://legacy.spa.aalto.fi/software/warp/ (last accessed: 11/08/2017)